\documentclass[letterpaper,twocolumn,10pt]{article}
\usepackage{usenix2019_v3}

\usepackage{tikz}
\usepackage{amsmath}

\usepackage{filecontents}
\usepackage{amsthm}
\newtheorem{myDef}{Definition} 

\begin{document}

\date{}

\title{\Large \bf HinDom: A Robust Malicious Domain Detection System based on Heterogeneous Information Network with Transductive Classification}

\author{
{\rm Xiaoqing Sun\textsuperscript{1}, Mingkai Tong\textsuperscript{2}, Jiahai Yang\textsuperscript{3}}\\
 Institute for Network Sciences and Cyberspace,\\ Tsinghua University\\
 National Research Center for Information Science \\and Technology, Beijing, China
\and
{\rm Xinran Liu\textsuperscript{4}, Heng Liu\textsuperscript{5}}\\
\textsuperscript{4} National Computer Network Emergency \\
Response Center, Beijing, China\\
\textsuperscript{5} Institute for Network Sciences \\ and Cyberspace Beijing, China
} 

\maketitle
\pagestyle{empty}
\thispagestyle{empty}
\widowpenalty 10000
\clubpenalty 10000
\begin{abstract}
Domain name system (DNS) is a crucial part of the Internet, yet has been widely exploited by cyber attackers. Apart from making static methods like blacklists or sinkholes infeasible, some weasel attackers can even bypass detection systems with machine learning based classifiers. As a solution to this problem, we propose a robust domain detection system named \textit{HinDom}. Instead of relying on manually selected features, HinDom models the DNS scene as a Heterogeneous Information Network (HIN) consist of clients, domains, IP addresses and their diverse relationships. Besides, the metapath-based transductive classification method enables HinDom to detect malicious domains with only a small fraction of labeled samples. So far as we know, this is the first work to apply HIN in DNS analysis. We build a prototype of HinDom and evaluate it in CERNET2 and TUNET. The results reveal that HinDom is accurate, robust and can identify previously unknown malicious domains.
\end{abstract}

\section{Introduction}
Though improved increasingly, the Internet is still widely used by adversaries who misuse benign services or protocols to run malicious activities. As a foundation of the Internet, Domain Name System (DNS) provides mappers among IP addresses and domain names, identifying services, devices or other resources in the network. As a consequence, domains are one of the major attack vectors used in various cybercrimes, such as spams, phishing, malware and botnets, etc. Therefore, it is essential to effectively detect and block malicious domains when combating cyber attackers.

After some flexibility-increasing techniques (e.g. Fast-Flux, Domain-Flux, Double-Flux, etc) make static block methods like blacklists infeasible, extensive researches are proposed for malicious domain detection. Traditional systems \cite{antonakakis2010building,bilge2014exposure,
antonakakis2012throw,antonakakis2011detecting,
liu2017don,chiba2016domainprofiler} mostly follow a feature based approach. Though these researches get relatively good performance, potential problems are commonly ignored. First, in the training phase, these detection systems require labeled datasets large enough to guarantee accuracy and coverage. However, the fickle nature of DNS makes accurate labeling an arduous process. Second, it seems they treat each domain individually and rely on some manually selected statistical features (e.g. number of distinct IP addresses, the standard deviation of TTL, etc), making the detection system easy to be evaded by sophisticated attackers\cite{geffner2013end,anderson2016deepdga, chen2017practical}. Some researchers \cite{manadhata2014detecting, rahbarinia2015segugio, khalil2016discovering, zou2015detecting} intend to utilize structural information for a more robust detection system. However, under the limitations of homogeneous network methods,  almost all these researchers model the DNS-related data into a client-domain bipartite graph\cite{rahbarinia2015segugio} or a domain-IP bipartite graph\cite{khalil2018domain}. In this case, they can represent at most two types of entities and utilize only one kind of relationship, leaving plenty of information untapped. 

Facing the problems mentioned above, we propose an intelligent domain detection system named \textbf{\textit{HinDom}}. First, to fuse more information and introduce higher-level semantics, we model the DNS scene into a Heterogeneous Information Network (HIN), as a HIN model can represent diverse components and relations. Second, a transductive classification method is applied to make use of the structural information, and therefore reduces the dependence on labeled datasets. Besides, considering real-world practicality, we design a series of filtering rules to improve efficiency and reduce noises.

In HinDom, we hold the intuitions that, 1) a domain which has strong associations with the known malicious domains is likely to be malicious and 2) attackers can falsify domains individually but cannot easily distort their associations. Thus, to be more robust against attackers' evasion tactics, we first naturally model the DNS scene into a HIN with client nodes, domain nodes, IP address nodes and the following six types of relations among them: (i) \textit{Client-query-Domain}, client \textit{a} queries domain \textit{b}. (ii) \textit{Client-segment-Client}, client \textit{a} and client \textit{b} belong to the same network segment. (iii) \textit{Domain-resolve-IP}, domain \textit{a} is resolved to IP address \textit{b}. (iv) \textit{Domain-similar-Domain}, domain \textit{a} and domain \textit{b} have similar character-level distribution. (v) \textit{Domain-cname-Domain}, domain \textit{a} and domain \textit{b} are in a CNAME record. (vi) \textit{IP-domain-IP}, IP address \textit{a} and IP address \textit{b} are once mapped to the same domain. Then multiple meta-paths are built to represent connections among domains and the PathSim algorithm \cite{sun2011pathsim} is applied to compute the similarity among domain nodes. The similarities derived from different meta-paths are combined according to Laplacian Scores\cite{he2006laplacian}, excavating associations among domains over multiple views. Finally, illuminated by LLGC\cite{zhou2004learning}, GNetMine\cite{ji2010graph} and HetPathMine\cite{luo2014hetpathmine}, a meta-path based transductive classification method is introduced to HinDom to make full use of the information provided by unlabeled samples.

To sum up, we make the following contributions in this
research:

1) A comprehensive represent model. We naturally represent the DNS scene by modeling clients, domains, IP addresses and their diverse relations into a HIN. To the best of our knowledge, this is the first work to introduce HIN in malicious domain detection. The combined domain similarity formulated over multiple meta-paths fully represents the rich semantics contained in DNS-related data.

2) Transductive classification in HIN. To reduce the cost of obtaining label information, we apply a meta-path based transductive classification method in HinDom. The experiment results show that HinDom yields ACC: 0.9960, F1-score: 0.9902 with 90\% labeled samples and can still detect malicious domains with ACC: 0.9626,  F1-score: 0.9116 when the initial labeled sample rate decreases to 10\%.

3) Practicality evaluation. We implement a prototype of HinDom and evaluate its performance in two realistic networks, CERNET2 and TUNET. During the deployment, we are able to detect long-buried mining botnets in these two educational networks. The evaluation results show that HinDom is practical in real-world and can identify malicious domains unrevealed by public blacklists.

The rest of this paper is organized as follows. After presenting related work in Section 2 and introducing necessary preliminaries in Section 3, we describe HinDom's framework and the technical details of each component in Section 4. Section 5 reports the experiment results and real-world evaluations. We discuss limitations and future work in Section 6 and summarize our work in Section 7.

\section{Related Work}

\textbf{Malicious domain detection.} As static block methods like blacklists become infeasible, plenty of researches have been proposed to detect malicious domains. We group them into two categories: \textit{object-based approaches} and \textit{association-based approaches}. It is hard to practice a fair comparison between HinDom and these prior researches, as both their datasets and system implementations are unavailable. In this section, we provide detailed introductions of researches in each group and discuss why HinDom is more advanced.

\textit{object-based approaches}. Their general method is to first build a classifier based on features extracted from various DNS-related data. Then after being trained with a ground truth dataset, the classifier can be used to inspect unlabeled domains. Notos\cite{antonakakis2010building} assigns reputation scores to domains by analyzing network and zone features. It trains classifiers to measure a domain's closeness with five pre-labeled groups (Popular, Common, Akamai, CDN and Dynamic DNS) and uses the calculated scores as features for final detection. Exposure\cite{bilge2014exposure} extends the scope of detection to malicious domains involved in spams, phishing, etc and obtains higher efficiency with lower requirements for training data. Kopis \cite{antonakakis2011detecting}  gets larger visibility by leveraging traffic among top-level-domain servers. Some works aim at detecting a specific kind of malicious domains. Pleiades \cite{antonakakis2012throw} detects algorithmically generated domains (AGDs) by analyzing NXDomain responses in DNS traffic while others \cite{schuppen2018fanci, woodbridge2016predicting, lison2017automatic} focus on AGDs' distinguish character distributions. In these researches, various resources are accessed for data enrichment (e.g. ASN, WHOIS, geo-location, network traffic, etc), yet they are analyzed in a coarse-grained way. The classifier treats each domain individually and relies on many statistic results as features, which makes the detection system easy to be evaded by sophisticated attackers. For instance, character patterns of malicious domains can be designed to imitate those of the benign ones\cite{geffner2013end,anderson2016deepdga}. It is also easy for attackers to change temporal patterns like request intervals or TTL values, which commonly service as major features of the classifiers. HinDom is more robust by further utilizing the rich structural information among domains.

\textit{association-based approaches}. Systems in this group get more macro perspectives by utilizing the relationships among domains. Manadhata et al.\cite{manadhata2014detecting} build a bipartite client-domain graph and apply belief propagation to discover malicious domains. Segugio\cite{rahbarinia2015segugio} focuses on the \textit{who is querying what} information and constructs a machine-domain bipartite graph based on DNS traffic between clients and the resolver. Khalil et al.\cite{khalil2016discovering} build a domain-IP graph based on a passive DNS dataset and then simplify it to a domain graph for detection. Futai Zou et al. \cite{zou2015detecting} try to utilize both the client-query-domain relation and the domain-resolve-IP relation
by constructing a DNS query response graph and a passive DNS graph. However, due to the limitations of homogeneous network analysis methods, all the above researches can represent at most two types of nodes and utilize only one type of relationship, leaving plenty of information in DNS-related data untapped. HinDom solves this problem with a HIN model which can represent multiple types of nodes and relations for a more comprehensive analysis.

\textbf{Heterogeneous information network}. In recent years, an increasing number of researches start to focus on the importance of heterogeneous information network and apply it to various fields, such as link prediction, recommender system, information fusion, etc \cite{shi2017survey}. Hindroid \cite{hou2017hindroid} is the first work to apply HIN in information security field. By analyzing different relations among API calls in Andriod program, HinDriod extracts higher-level semantics to discover Android malware precisely. Scorpion \cite{fan2018gotcha} use HIN to model relations among archives, files, APIs and DLLs for malware detection. As for transductive classification in HIN, GNetMine \cite{ji2010graph} is the first work to expand a transductive classification method named LLGC\cite{zhou2004learning} from homogeneous network to HIN. HetPathMine \cite{luo2014hetpathmine} utilizes metapaths to set different classification criterions for different types of objects. Grempt \cite{wan2015graph} generates local estimated labels for unlabeled samples and expands the transductive method from classification to regression. Illuminated by these researches, our work shows HIN's usefulness in malicious domain detection.

\section{PRELIMINARIES}

\subsection{Heterogeneous Information Network}
In the real world, most systems contain diverse interactions among different types of components. However, for ease of analysis, they are usually modeled as homogeneous networks with unique type of nodes and links. In this case, information loss is caused by ignoring differences among objects and relationships. Recently, researchers start to model these systems into Hetergeneous Information Networks (HINs)\cite{sun2009ranking}, which can fuse richer semantics and support more comprehensive represents. The basic concepts of HIN are as follows.

\begin{myDef}
\textbf{Hetergeneous Information Network (HIN)} \cite{sun2009ranking}. Given a graph G=$\langle V,E \rangle$, where V is the set of nodes, E is the set of links. m types of objects are denoted as $V_1$=$\lbrace v_{11},v_{12},...,v_{1n_1}\rbrace,...,V_m$=$\lbrace v_{m1},v_{m2},...v_{mn_m}\rbrace$, where $n_i$ is the number of the $i$-th type nodes and p types of relationships are denoted as $E_1$=$\lbrace E_{11},E_{12},$...$,E_{1q_1}\rbrace$,...,$E_p$=$\lbrace e_{p1},e_{p2},...e_{pq_p}\rbrace$, where $q_i$ is the number of the $i$-th type of relations. We regard G as a HIN if $m\geq 2$ or $p\geq 2$. When m=p=1, G reduces to a homogeneous network.
\end{myDef}

\begin{myDef}
   \textbf{Network Schema}\cite{sun2009ranking}. $T_G=\langle A,R\rangle$ is the network schema of a HIN $G=\langle V,E\rangle$, with type mapping function $\varphi: V\rightarrow A$ and $\psi: E\rightarrow R$, where A is the set of object types and R is the set of relationship types.
\end{myDef}

\begin{myDef}
   \textbf{metapath}\cite{sun2011pathsim}. Given a network schema $T_G=\langle A,R\rangle$, a metapath $P$ defines a composite relation $R=R_1\circ R_2\circ ...\circ R_L$ between $A_1$ and $A_{L+1}$, where $\circ$ is the relation composition operator. $P$ is denoted as $A_1\xrightarrow{R_1} A_2\xrightarrow{R_2}...\xrightarrow{R_L} A_{L+1}$, where $L$ is the length of the metapath.
\end{myDef}

\begin{figure}[hbtp]
	\centering
	\includegraphics[width =0.5\textwidth]{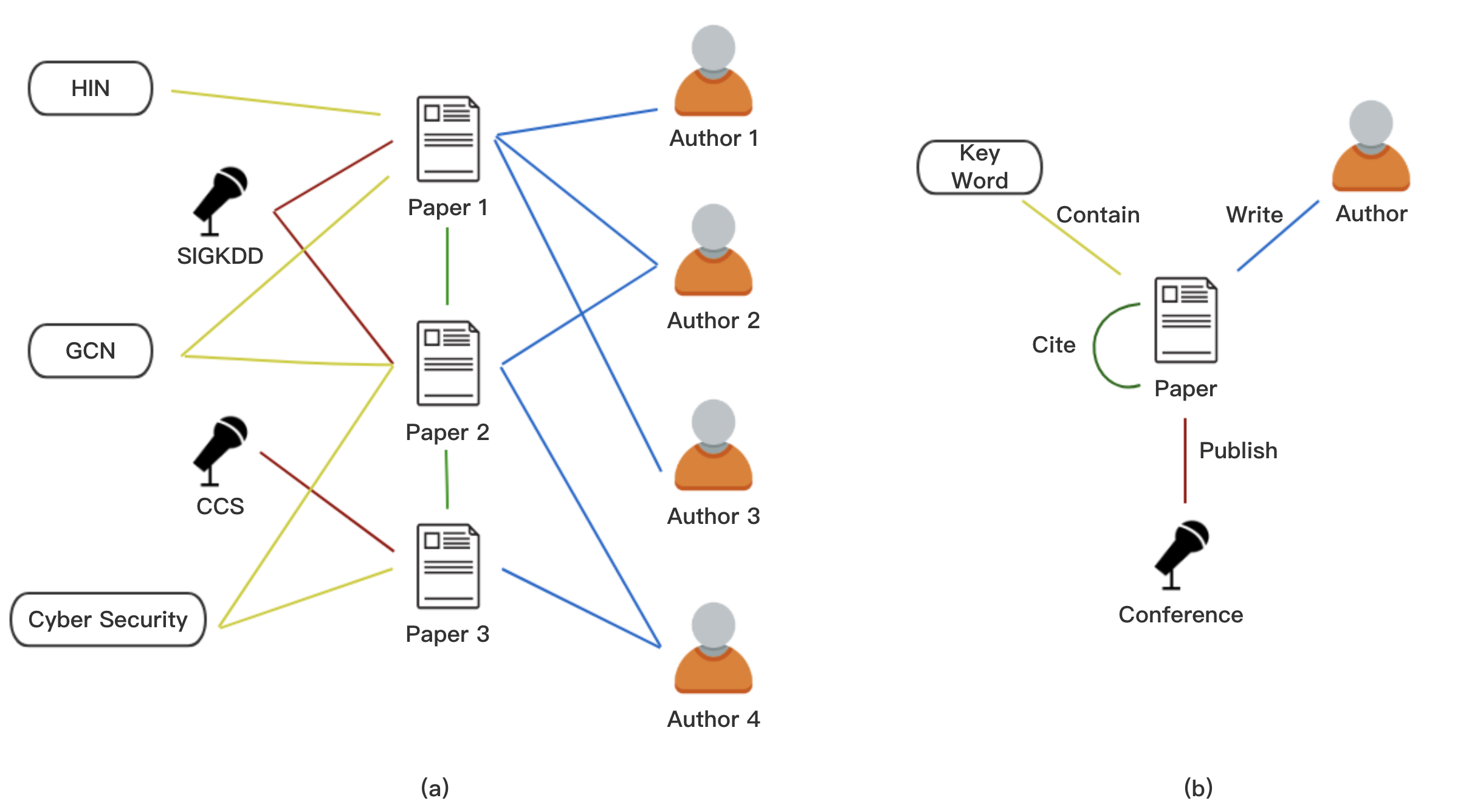}	
	\caption{An example of HIN instance (a) and its network schema (b)}
\end{figure}

Figure 1 shows a HIN model of the bibliographic dataset DBLP\cite{DBLP}. It represents four types of nodes: paper (P), author (A), conference (C) and keyword (K), as well as four kinds of links: authors write papers, papers are published in conferences, papers contain keywords and a paper cites other papers. The relation between author and conference can be represented as metapath A $\xrightarrow{write} P\xrightarrow{published} $ C, or $APC$ for short. Metapath APCPA indicates the relation that authors have published papers on the same conference. To sum up, a HIN instance contains detailed information while its network schema describes the structural constraints and metapaths are used to represent complex relations among entities.

\subsection{Transductive Classification}
Unlike inductive classification, instead of learning general decision functions from training data, transductive classification infers from specific training cases to specific test cases. The situation is more like to propagate label information over the whole network. Therefore, when there are many test samples but few labeled training samples, transductive methods can classify more effectively with the utilization of information from the unlabeled data. Based on definitions in Section 3.1, transductive classification in HIN can be defined as follow.
\begin{myDef}
\textbf{Transductive classification in HIN}\cite{ji2010graph}. Given a HIN $G=\langle V,E\rangle$ and a subset of its labeled nodes $\tilde{V} \subseteq V$ with their label information denoted by vector Y, transductive classification is to predict labels for nodes in $V-\tilde{V}$.
\end{myDef}


\section{HinDom System Description}
\begin{figure*}[hbtp]
	\centering
	\includegraphics[width =\textwidth]{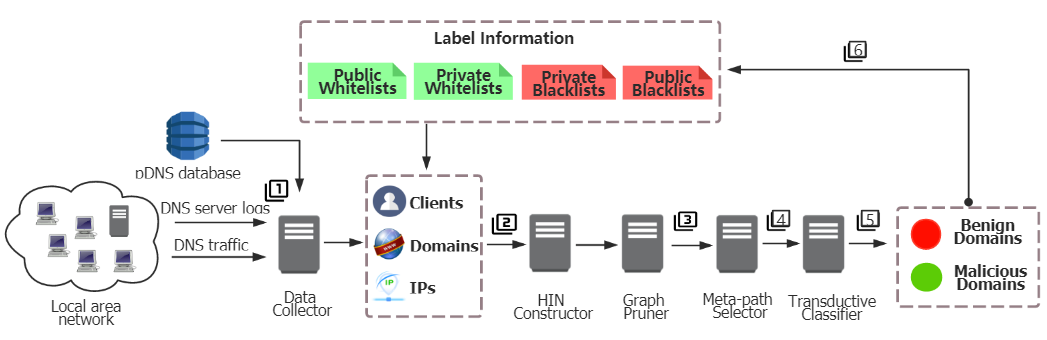}	
	\caption{The architecture of HinDom}
\end{figure*}

The intuition of HinDom is that domains with strong relationships tend to belong to the same class (benign or malicious).  Besides, attackers can only falsify domain's features individually but cannot easily control the natural associations generated in DNS. We model clients, domains, IP addresses as well as their relations into a HIN and analyze six types of associations among domains based on the following two observations: (i) Attackers are subjected to the cost of network resources. That is, though trying to stay dynamic, attackers tend to reuse network resources due to economic constraints. (ii) The set of malicious domains queried by victims of the same attacker tend to overlap.

As shown in Figure 2, HinDom has five main components: Data Collector, HIN Constructor, Graph Pruner, Meta-path Combiner and Transductive Classifier. After DNS-related data are collected (step 1), a HIN consist of clients, domains, IP addresses and their various relations is constructed to represent the DNS scene (step 2). Then some nodes in the graph are pruned to filter noises and reduce computing complexity (step 3). We analyze six different meta-paths and combine them according to their influences on domain detection (step 4). Finally, based on some initial label information, the transductive classifier categorizes unlabeled domains (step 5). We analyze the classification result and add it to private whitelist or blacklist for further detection (step 6). In the following, we will introduce each component in detail.

\subsection{Data Collector}
To obtain richer information that reveals the behavior of actual users, instead of sending specific DNS queries on purpose, we execute DNS data collection passively. Three major data sources that we collect are: (i) \textit{DNS server log}. When dealing with queries, DNS servers generate logs to collect information like source IP, queried domain, time, etc. Among all the logs, those of the recursive servers are widely used to extract information about "who is querying what" in local area networks. (ii) \textit{DNS traffic}. It contains the most comprehensive information with various fields such as NS, MX, TXT, PTR, etc. Yet, considering privacy issues, this kind of data is hard to share publicly. (iii) \textit{Passive DNS dataset}. Some organizations (e.g. Internet Systems Consortium, Farsight Security\cite{DNSDB}, 360 NetLab\cite{passivedns}, etc.) have constructed passive DNS (pDNS) systems with sensors voluntarily deployed by contributors in their infrastructures. They aggregate the captured DNS messages before making them publicly available. Records in pDNS do not contain client information. They only offer the first and last timestamps of a domain's appearance, as well as the total number of domain-IP resolutions in between.

The data collector collects resolver's logs or DNS traffic between clients and the resolver in a local area network (LAN) during a time window \textit{T}, which can be set to an hour, a day or a week, considering computing resources and the network size. Noting that HinDom can construct the HIN model just based on DNS response traffic. But pDNS dataset can provide richer information on domain-IP relations in both spatial and temporal dimensions. Besides, when DNS traffic data is unavailable due to permission or technique restrictions, HinDom can utilize DNS logs construct the client-domain part and use pDNS dataset for the domain-IP part.

\subsection{HIN Constructor}
As shown in Figure 3, based on the collected data, HinDom naturally models the DNS scene as a HIN consist of clients, domains, IP addresses and six types of relations among them. The details of these relations are as follows and Table 1 lists their corresponding adjacent matrices.

\begin{figure}[hbtp]
	\centering
	\includegraphics[width =0.5\textwidth]{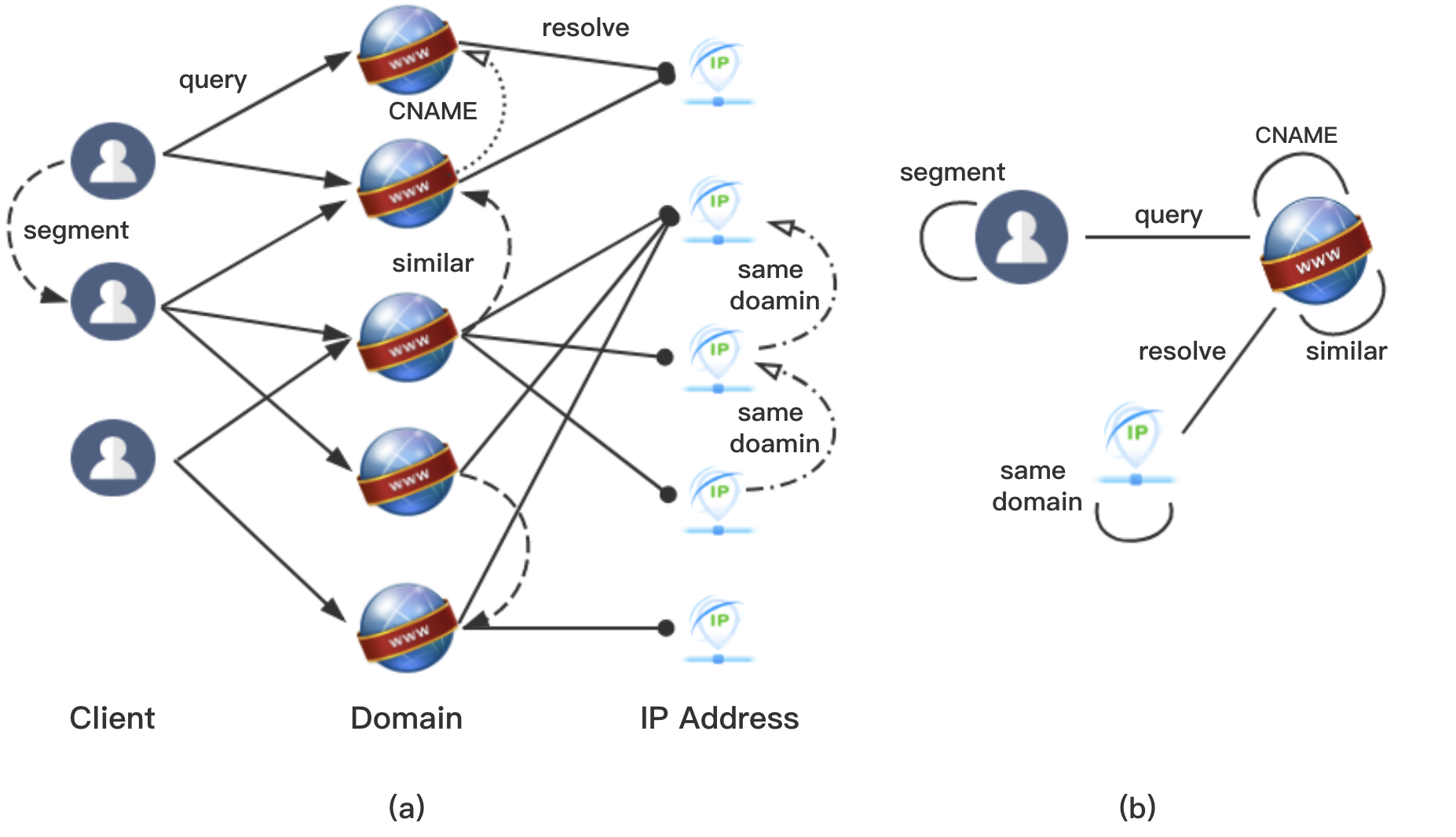}	
	\caption{HIN instance (a) and its network schema (b) in HinDom}
\end{figure}

\begin{itemize}
	\setlength{\itemsep}{0pt}
	\setlength{\parsep}{0pt}
	\setlength{\parskip}{0pt}
	\item \textbf{\textit{Client-query-Domain}}, we use matrix $Q$ to denote that domain $i$ is queried by client $j$.
	\item \textbf{\textit{Client-segment-Client}}, we use matrix $N$ to denote that client $i$ and client $j$ belong to the same network segment.
	\item \textbf{\textit{Domain-resolve-IP}}, we use matrix $R$ to denote that domain $i$ is resolved to IP address $j$.
	\item \textbf{\textit{Domain-similar-Domain}}, we use matrix $S$ to denote the character-level similarity between domain $i$ and $j$.
	\item \textbf{\textit{Domain-cname-Domain}}, we use matrix $C$ to denote that domain $i$ and domain $j$ are in a CNAME record.
	\item \textbf{\textit{IP-domain-IP}}, we use matrix $D$ to denote that IP address $i$ and IP address $j$ are once mapped to the same domain.
\end{itemize}

All these adjacent matrices can be naturally extracted from the DNS-related data except matrix $S$ which indicates the character-level similarity among domains. We use n-gram to process the domain name strings, regard the results of the entire dataset as a vocabulary and embed each domain into a characteristic vector. Then we use the K-Means algorithm to cluster these vectors into $K$ categories and transform the clustering result into matrix $S$. In our experiments, we test uni-gram, bi-gram and tri-grams for both types of features. Tri-grams brings a marginal improvement with much more cost on memory requirements. Considering performance and complexity, we concatenate uni-grams and bi-grams as features and empirically set K = 20.

\begin{table}[htbp]
\caption{Elements and descriptions of the relation matrices}
\begin{center}
\begin{tabular}{c c p{5.2cm}}
\hline
Matrix & Element & Description \\
\hline
Q & $q_{i,j}$ & if domain \textit{i} is queried by client \textit{j}, then $q_{i,j}$ = $1$, otherwise, $q_{i,j}$ = $0$.\\
N & $n_{i,j}$ & symmetric, if client \textit{i} and \textit{j} belong to the same network segment, then $n_{i,j}$ = $n_{j,i}$ = $1$, otherwise, $n_{i,j}$ =$n_{j,i}$ = $0$.\\
R&$r_{i,j}$ & if domain \textit{i} is resolved to ip \textit{j}, then $r_{i,j}$ = $1$, otherwise, $r_{i,j}$ = $0$.\\
S & $s_{i,j}$ & symmetric, if domain \textit{i} and \textit{j} are similar on character level, then $s_{i,j}$ = $s_{j,i}$ = $1$, otherwise, $s_{i,j}$ = $s_{j,i}$ = $0$. \\
C & $c_{i,j}$ & symmetric, if domain \textit{i} is the cname of domain \textit{j}, then $c_{i,j}$ = $c_{j,i}$ = $1$, otherwise, $c_{i,j}$ = $c_{j,i}$ = $0$.\\
D &$d_{i,j}$ & symmetric, if IP address \textit{i} and \textit{j} are once resolved to the same domain, then $d_{i,j}$ = $d_{j,i}$ = $1$, otherwise, $d_{i,j}$ = $d_{j,i}$ = $0$.
\end{tabular}
\label{tab1}
\end{center}
\end{table}
\subsection{Graph Pruner}
Because we aim at detecting malicious domains in a campus or enterprise network, the HIN may contain millions of nodes and billions of edges, it is a waste of computing resources to model all these entities and perform the corresponding matrix operations.  Besides, The data we collect directly from DNS traffic or logs is quite dirty with noises like irregular domains, "large" clients, etc. Therefore, we add a graph pruning module in HinDom to improve its performance and practicality. The graph pruner filters nodes according to the following conservative rules.

\begin{itemize}
	\setlength{\itemsep}{0pt}
	\setlength{\parsep}{0pt}
	\setlength{\parskip}{0pt}
	\item \textbf{Unusual domains}. We remove domains that fail to meet the naming rules, for example,  \textit{icmsb2018(at)163.com}, which may be caused by mistyping, misconfiguration or benign services' misuse. Besides we discard domains that are queried by only one client to focus on those that have greater impacts over the LAN.
	\item \textbf{Popular domains}. In most cases, popular domains queried by a large fraction of clients in a LAN are inclined to be benign; otherwise, there will be a significant attack event and will be easily detected by the security management department. Besides, these popular domains cause much computational complexity as they are all nodes with high degrees in HIN. Therefore, we filter out domains that queried by $K_d$\% clients in a network. To be conservative, we set $K_d$\% to be 25\% in our experiments.
	\item \textbf{Large clients}. There are some "large" clients outstanding by querying a large fraction of the whole domain set. We find these devices are often DNS forwarders or large proxies and thus can not represent the behavior of regular clients. HinDom removes them to eliminate the ambiguousness and complicacy they bring into the system. In our evaluations, the top $K_a$\% (empirically set to 0.1\%) most active clients are discarded.
	\item \textbf{Inactive clients}. We regard clients that query less than $K_c$ domains as the inactive ones. They are discarded for the lack of effects on mining associations among domains. In our experiments, $K_c$ is set to be 3.
	\item \textbf{Rare IPs}. For the same reason as above, IP addresses that only map to one domain are also filtered out to boost performance and save computing resources.
\end{itemize}
To be more conservative for information loss caused by graph pruning, we set some exceptions against the mentioned rules based on label information. Considering that some attackers try to hide by reducing activities, we keep domains with clear malicious labels even if they are regarded as unusual ones. Same to their related clients or IP addresses.

\begin{table*}[htbp]
\caption{Commuting Matrix of each metapath}
\begin{center}
\begin{tabular}{c c c p{8cm}}
\hline
PID& metapath& Commuting Matrix M& Description \\
\hline
1 & $d \xrightarrow{S} d$& S& domains similar to each other on character level\\
2 & $d \xrightarrow{C} d$ &C & the cname relationship among domains\\
3 &  $d \xrightarrow{Q} c \xrightarrow{Q^T}  d$ & $QQ^T$ & domains queried by same clients\\
4&$d \xrightarrow{R} ip \xrightarrow{R^T}  d$ & $RR^T$& domains resolved to same IP address \\
5 & $d \xrightarrow{Q} c \xrightarrow{N} c\xrightarrow{Q^T} d$&$QNQ^T$& domains queried by clients belong to the same subnet\\
6 & $d \xrightarrow{R} ip \xrightarrow{D} ip\xrightarrow{R^T} d$&$RDR^T$& domains resolved to IPs that belong to the same attacker\\
\end{tabular}
\label{tab1}
\end{center}
\end{table*}

\subsection{Meta-path Combiner}
As mentioned in Section 3, meta-paths are used in HIN to denote complex associations among nodes. Because we are interested in malicious domain detection, HinDom only chooses symmetric meta-path where $A_1=A_{L+1}=\textit{domains}$ and derives six types of meta-paths from the six relations mentioned above. Table 2 displays the description and corresponding commuting matrix $M_k$ of each meta-path while the reasons for choosing them are listed as follows.

\begin{itemize}
	\setlength{\itemsep}{0pt}
	\setlength{\parsep}{0pt}
	\setlength{\parskip}{0pt}
	\item \textbf{P1: $d \xrightarrow{S} d$}. We have noticed that benign and malicious domains differ in character distributions. Besides, malicious domain names from the same family tend to follow a similar textual pattern.
	\item \textbf{P2: $d \xrightarrow{C} d$}. The cname domain of a benign domain is unlikely to be malicious, vice versa.
	\item \textbf{P3: $d \xrightarrow{Q} c \xrightarrow{Q^T}  d$}. Infected clients of the same attackers tend to query partially overlapping sets of malicious domains while normal clients have no reasons to reach out for them.
	\item \textbf{P4: $d \xrightarrow{R} ip \xrightarrow{R^T}  d$}. IP resources are relatively stable in Internet, domains resolved to the same IP address in a period tend to belong to the same class.
	\item \textbf{P5: $d \xrightarrow{Q} c \xrightarrow{N} c\xrightarrow{Q^T} d$}. Adjacent clients are vulnerable to the same attacks. For example, malware propagating in a subnet or spams aiming to clients on the same segment.
	\item \textbf{P6: $d \xrightarrow{R} ip \xrightarrow{D} ip\xrightarrow{R^T} d$}. Even trying to keep flexible, with funding limits, attackers are likely to reuse their domain or IP resources.

\end{itemize}

Based on meta-paths, an algorithm named PathSim\cite{sun2011pathsim} can be used to measure the similarity among nodes. Yet different meta-paths represent associations from different points of view which are not equally important in malicious domain detection. HinDom obtains a combined meta-path with the corresponding similarity matrix denoted as follow, where $w_k$ is the weight assigned to each meta-path.
$$
M'=\sum_{k=1}^{6}\omega_k \cdot PathSim(M_k)=\sum_{k=1}^{6}\omega_k \cdot \dfrac{2M_{k(i,j)}}{M_{k(i,i)}+M_{k(j,j)}}
$$

Many methods can be used to compute the weight vector, for example, linear regression with gradient descent. HinDom chooses to use the Laplacian Score (LS)\cite{he2006laplacian} for two reasons: First, LS can be applied to unsupervised situations. Second, as a "filter" method, LS is independent of further learning algorithms and can evaluate features directly from the local geometric structure of data. The basic idea of LS is to evaluate features according to their locality preserving power. LS constructs a nearest neighbor graph and seeks features respecting this graph structure. Specifically, We code all meta-paths into a tensor $T\in R^{6\times n\times n}$, where $T_{k,i,j}=M_k(i,j)$, $n$ is the number of domains, $M_k$ is the commuting matrix of meta-path $P_k$. Then a domain meta-path representation matrix $W\in R^{n\times m}$, where $W_{k,i}=\sum_j T_{k,i,j}$ is generated as the input of LS.

\subsection{Transductive Classifier}

Though some public domain lists are commonly used as label information in malicious domain detections, some subtle issues are ignored. For whitelists, the widely used Alexa top K list only contains second-level domains (2LD) sorted by popularity, which leads to many false positives. For example, a prevalent 2LD may hold proxies to malicious activities and some malicious domains may rank high with a burst of queries from the infected clients. As for blacklists, though usually generated with robust evidences, some discrepancies are still caused by the fickle nature of DNS. For instance, domains like \textit{alipay.com} are in DGArchive \cite{plohmann2016comprehensive}, a database of DGAs and the corresponding domains. Besides, when new malicious domains come, blacklists cannot update in time. In a word, none of these lists is completely reliable. It is a time-consuming and cost-expensive process to obtain an accurately labeled dataset as the ground truth.

To reduce the cost of labeling, HinDom applies a meta-path based transductive classification method which can perform well even with a small fraction of labeled samples. The basic two assumptions in transductive classification are (i) \textit{smoothness assumption}, objects with tight relationships tend to belong to the same class; (ii) \textit{fitting assumption}, the classification results of the known nodes should consist with the pre-labeled information. Therefore, the cost function of the transductive classifier is as follow,

$$
Q(F)=\dfrac{1}{2}(\sum_{i,j=0}^{n-1}M'_{i,j}\Vert\dfrac{F_i}{\sqrt{D_{ii}}}-\dfrac{F_j}{\sqrt{D_{jj}}}\Vert^2+\mu\sum_{i=0}^{n-1}\Vert F_i-Y_i\Vert^2)
$$
where $n$ is the number of domain nodes in HIN, $M'\in R^{n\times n}$ is the similarity matrix we get from the combined metapath, $D\in R^{n\times n}$ is a diagonal matrix whose (\textit{i, i})-element equals to the sum of the \textit{i}-th row of $M'$. $F \in R^{n\times 2}$ contains each domain's probability of being benign or malicious while $Y\in R^{n\times 2}$ denotes their pre-labeled information. We can see the first term of this cost function represents the smoothness assumption while the second term follows the fitness assumption. Trade-off between the two assumptions is adjusted by parameter $\mu$.
In order to find the $F^*$ that minimize $Q(F)$, we get
$$
\dfrac{dQ}{dF}=F^*-F^*S+\mu(F^*-Y)=0
$$
$$
F^*=\beta(I-\alpha S)^{-1}Y
$$
where $\alpha=\dfrac{1}{1+\mu}$, $\beta=\dfrac{\mu}{1+\mu}$ and $S=D^{-1/2}M'D^{-1/2}$.

We get the theoretical optimal solution, yet in the real world, inverting a large matrix will consume too much computing resources. Thus, illuminated by LLGC\cite{zhou2004learning}, in HinDom we perform iterations $F(t+1)=\alpha SF(t)+\beta Y$ to approach the optimal solution. We refer the readers to LLGC\cite{zhou2004learning} for the theoretical proof that this iteration can coverage to the optimal solution.
The algorithm of Transductive Classifier is summarized as follow.

\textbf{Step 1}, Given a HIN $G=\langle V,E\rangle$ with incomplete domain label information $Y$, get similarity matrix $M'$ from Metapath Combiner.

\textbf{Step 2}, Regularize the similarity matrix with $S=D^{-1/2}M'D^{-1/2}$, where $D$ is a diagonal matrix whose (\textit{i, i}) - element equals to the sum of the \textit{i}-th row of $M'$.

\textbf{Step 3}, Set $F(0)=Y$, iterate $F(t+1)=\alpha SF(t)+\beta Y$ until it converges.

\textbf{Step 4}, Label domain $i$ `\textit{benign}' if $F_{i,0}\geq F_{i,1}$, vice versa.

We further analyze the classification results and add domains with solid labels, namely the difference between $ F_t[i,0]$ and $F_t[i,1]$ is higher than the threshold $\theta$, into local whitelist or blacklist as a supplement for further detection. Considering the dynamic nature of DNS, we only keep local label information within 7 days.

\section{Experiments}

In this section, we present comprehensive experiments to evaluate HinDom from three aspects: performance, robustness and practicality. For performance, we first analyze detection results and the corresponding weight of each meta-path to prove the effectiveness of Meta-path Combiner. Then, we test HinDom in insufficient labeling scenario and multi-classification scenario. For robustness, we test HinDom's ability to deal with label noises in the training dataset. For practicality, we test HinDom when only public labels are available and deploy it in two real-world networks: CERNET2 and TUNET.

\subsection{Setup}

We evaluate HinDom in two real-world networks: CERNET2 and TUNET. Our research has obtained permissions from the relevant security management teams. The DNS-related data we get has been processed to minimize privacy disclosure, for example, the IP addresses of clients are desensitized by numerical identifiers.

\textbf{CERNET2}, the second generation of China Education and Research Computer Network. Jointly built by 26 universities, CERNET2 is the first IPv6 national backbone network in China and is the world’s largest next-generation Internet backbone network using pure IPv6 technology. At present, CERNET2 has 25 core nodes distributed in 20 cites with $2.5G\sim10Gbps$ bandwidth and provides IPv6 access services for more than 5 million users in about 500 research institutes. We capture DNS traffic in CERNET2 at Tsinghua node.

\textbf{TUNET}, the campus network of Tsinghua University. By statistics, we find that over 0.24 million clients request about 1.5 million unique domains per day. With close supervision and control, TUNET is much purer than CERNET2 and hides less malicious domains. 

In this research, we use DNS traffic of CERNET2 to construct its HIN and just use 360 pDNS dataset for data enrichment. As for TUNET, due to permission restrictions, we only get the logs of its central DNS resolver. The logs and 360 pDNS dataset are used respectively to construct HIN's domain-client part and domain-IP part. Besides, as for DNS traffic, we only use A, AAAA and CNAME records currently and may expand to PDG, MX, SRV, NS, PTR for richer information in the future.

\begin{table}[htbp]
\small
\caption{Description of the testing HIN instance}
\begin{center}
\begin{tabular}{|c|c|c|c|c|c|}
\hline
\multicolumn{6}{|c|}{\textbf{Nodes}}\\
\hline
Clients&\multicolumn{2}{|c|}{Benign Domains}& \multicolumn{2}{|c|}{Malicious Domains}&IPs  \\
\hline
\textasciitilde 0.49M&\multicolumn{2}{|c|}{\textasciitilde 0.7M}&\multicolumn{2}{|c|}{\textasciitilde 0.25M}&\textasciitilde 0.26M\\
\hline
\multicolumn{6}{|c|}{\textbf{Edges} (C-clients,D-domains, IP-IPs,c-cname,s-similar)}\\
\hline
C-C&C-D&D-c-D&D-s-D&D-IP&IP-IP\\
\hline
\textasciitilde 93M&\textasciitilde 112M&\textasciitilde 1.3M&\textasciitilde 15M &\textasciitilde 3.1M& \textasciitilde 4.3M\\
\hline
\multicolumn{3}{|c|}{\textbf{Nodes Total}}&\multicolumn{3}{|c|}{\textbf{Edges Total}}\\
\hline
\multicolumn{3}{|c|}{\textasciitilde 1.7M}&\multicolumn{3}{|c|}{\textasciitilde 228.7M}\\
\hline
\end{tabular}
\label{tab1}
\end{center}
\end{table}

To build the test dataset, we labeled about 1 million domains queried in CERNET2 on 13 April 2018 by referring to various whitelists/blacklists and expertise. For benign information, we regard domains whose 2LD appear in Alexa Top 1K list \cite{Amazon} or our local whitelist as benign ones. For malicious domains, we use multiple sources like \textit{Malwaredomains.com}\cite{Malwaredomains}, \textit{Malwaredomainlist.com}\cite{Malwaredomainlist}, DGArchive\cite{plohmann2016comprehensive}, etc. We also refer to integrated services like
Google Safe Browsing\cite{Google_Safe_Browsing} and VirusTotal\cite{Virustotal}. Besides, we manually investigate dubious domains which have both benign and malicious information. According to the graph pruning rules mentioned in Section 4.3, we discard some unhelpful nodes by setting $K_d$ = 25, $K_a$ = 0.1, $K_c$ = 3. Table 3 shows details about the HIN instance we construct and Table 4 lists the evaluation metrics used in the experiments.
\begin{table}[htbp]
\small 
\caption{Metrics for evaluation}
\begin{center}
\begin{tabular}{c p{6.5cm}}
\hline
Metric & Description \\
\hline
TP & malicious domains labeled as malicious \\
FP & benign domains labeled as malicious\\
TN & benign domains labeled as benign\\
FN & malicious domains labeled as benign\\
accuracy & $(TP +TN)/(TP + FP + TN+FN)$\\
precision &$TP/(TP + FP)$\\
recall & $TP/(TP + FN)$\\
F1 & $2\times (precision \cdot recall)/(precision + recall)$\\
ROC &a curve plotting TPR against FPR with various thresholds\\
AUC &area under the ROC curve\\
\end{tabular}
\label{tab1}
\end{center}
\end{table}

\subsection{Metepath Combiner}

HinDom relies on six meta-paths to formulate the final similarity matrix. These meta-paths represent associations among domains from different perspectives and thus have different influences on domain detection. We test the detection performance of each meta-path by randomly keep the label information for 70\% domains, leaving the remained 30\% as test samples and repeat the procedure for 10 times. The accuracy, precision and recall of labels generated by each meta-path are shown in Table 6 while the ROCs are shown in Figure 4.

\begin{table}[htbp]
\small
\caption{Labeled Metrics of each meta-path}
\begin{center}
\begin{tabular}{|c|c|c|c|c|}
\hline
\textbf{PID}&\textbf{metapath}&\textbf{accuarcy}&\textbf{precision}&\textbf{recall}\\
\hline
1&S&0.9376&0.8703&0.8712\\
\hline
2&C&0.9999&0.9999&0.9998\\
\hline
3&Q$Q^T$&0.9567&0.9198&0.8860\\
\hline
4&R$R^T$&0.9888&0.9879&0.9989\\
\hline
5&QN$Q^T$&0.9571&0.9251&0.8710\\
\hline
6&RD$R^T$&0.9754&0.9580&0.9281\\
\hline
\end{tabular}
\label{tab1}
\end{center}
\end{table}

\begin{figure}[hbtp]
	\centering
	\includegraphics[width =0.5\textwidth]{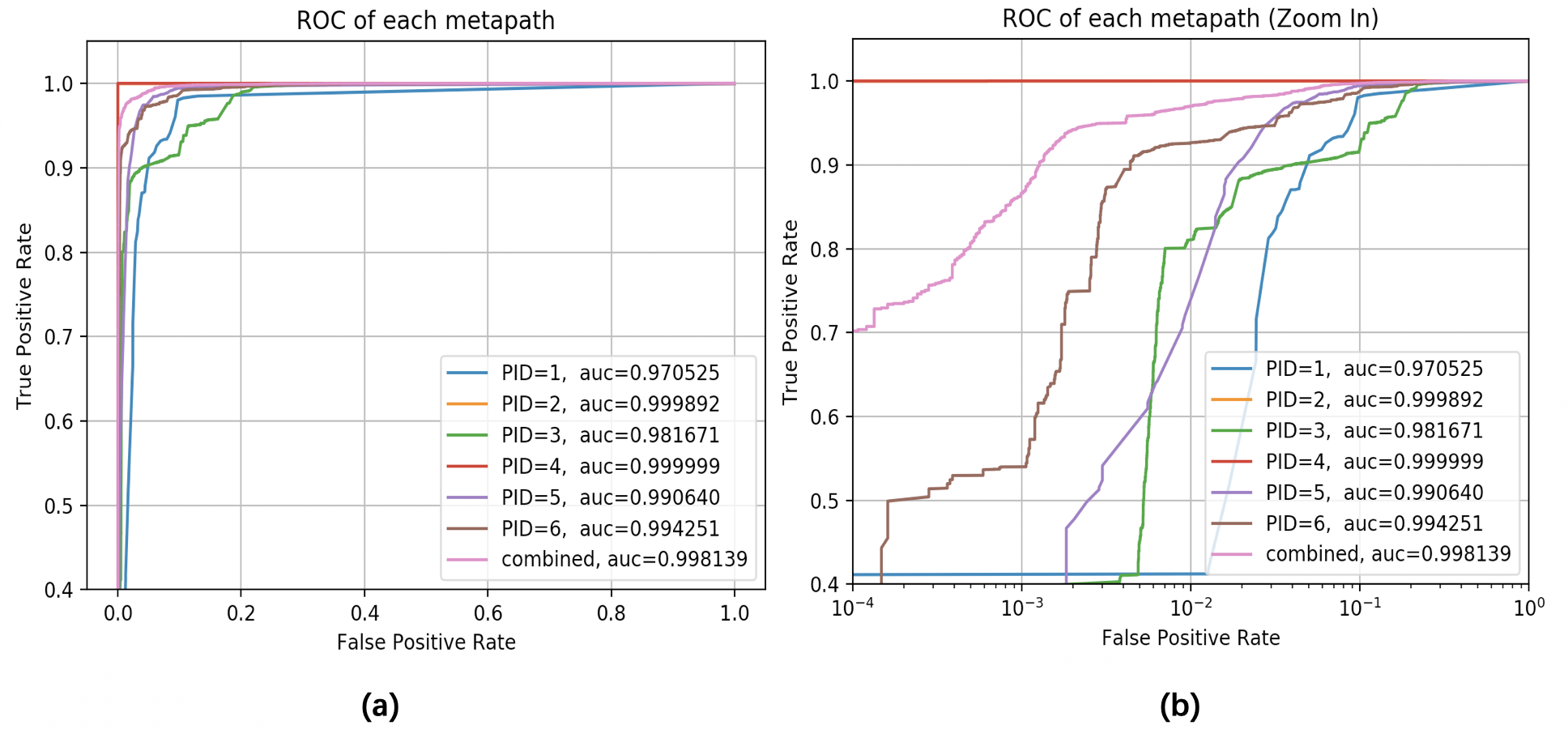}	
	\caption{ROC for the labeled result of each metapath}
\end{figure}

Noting that some meta-paths have low connectivity, which means the classifier cannot reach plenty of domains if it only relies on one of the meta-paths. Thus, when assigning weights to each meta-path, we need to consider two aspects: \textit{coverage} and \textit{accuracy}. For coverage, it means with this meta-path, we can fully exploit connections so that few domains will be left unlabeled. For accuracy, it means among the labeled domains, few are misclassified. The Meta-path Combiner assigns a weight to each meta-path according to Laplacian Scores. To test its effectiveness, we list the average detection results and the corresponding weight of each meta-path in Table 6, where the \textit{Unlabeled rate} and \textit{F1 score} reveal coverage and accuracy respectively. 

\begin{table}[htbp]
\small
\caption{Detection results of each meta-path}
\begin{center}
\begin{tabular}{|c|c|c|c|c|}
\hline
\textbf{PID}&\textbf{metapath}&\textbf{F1 Score}&\textbf{Unlabeled Rate}&\textbf{Weight}\\
\hline
1&S&0.8708&0.0027&0.1698\\
\hline
2&C&0.9996&0.9133&0.0003\\
\hline
3&Q$Q^T$&0.9026&0.0917&0.1386\\
\hline
4&R$R^T$&0.9934&0.5317&0.0125\\
\hline
5&QN$Q^T$&0.8973&0.0049&0.3826\\
\hline
6&RD$R^T$&0.9428&0.2057&0.2962\\
\hline
\multicolumn{2}{|c|}{\textbf{combined path}}&0.9743&0&-\\
\hline
\end{tabular}
\label{tab1}
\end{center}
\end{table}

\begin{table*}[htbp]
\caption{Detection results with different fraction of labels}
\begin{center}
\begin{tabular}{|c|c|c|c|c|c|c|c|c|}
\hline
\textbf{Initial Label}&\multicolumn{8}{|c|}{Metrics of each method}\\
\cline{2-9}
\textbf{Fraction}&\multicolumn{2}{|c|}{NB}&\multicolumn{2}{|c|}{SVM}&\multicolumn{2}{|c|}{RF}&\multicolumn{2}{|c|}{\textbf{HinDom}}\\
\cline{2-9}
&\textbf{accuarcy}&\textbf{F1 Score}&\textbf{accuarcy}&\textbf{F1 Score}&\textbf{accuarcy}&\textbf{F1 Score}&\textbf{accuarcy}&\textbf{F1 Score}\\
\hline
90\%&0.9632&0.9499&0.9864&0.9827&0.9813&0.9803&0.9960
&0.9905\\
\hline
70\%&0.9429&0.9276&0.9682&0.9550&0.9700&0.9611& 0.9880&0.9743\\
\hline
50\%&0.9020&0.8912&0.9286&0.9090&0.9257&0.9180&0.9840
&0.9776\\
\hline
30\%&0.8235&0.8260&0.8527&0.8516&0.8613&0.8544&0.9698
&0.9453\\
\hline
10\%&0.7929&0.7834&0.8120&0.8031&0.8141&0.7902&0.9626
&0.9116\\
\hline
\end{tabular}
\label{tab1}
\end{center}
\end{table*}

From Table 5, we can see that HinDom combines these meta-paths in order: PID 5, PID 6, PID 1, PID 3, PID 4 and PID 2. The combined path gets the ability to cover the whole set of domains with high detection accuracy. It is worth noting that some meta-paths with strong relations in domain detection (e.g. PID2: $d \xrightarrow{C} d$ ) get extremely high accuracy but very low coverage. This is consistent with the fact that domains in a CNAME record tend to belong to the same class, yet not many domains are in this kind of relationship. With Laplacian Score, the Meta-path Combiner assigns these meta-paths with relatively low weights to ensure that HinDom can detect as many domain names as possible. Take PID 4: $d \xrightarrow{R} ip \xrightarrow{R^T}  d$ and PID 6:$d \xrightarrow{R} ip \xrightarrow{D} ip\xrightarrow{R^T} d$ for instance, the latter extends its coverage by utilizing relations among IP addresses, though introducing some noises, it can reach more domains and thus plays a more important role in this scenario.

\subsection{Transductive Classification}

As mentioned in Section 3.5, the public whitelists or blacklists of domains are not completely reliable an have a number of subtle issues. In order to reduce the cost of labeling domains manually, HinDom utilizes a meta-path based transductive classification method to make better use of structural information of the unlabeled samples.  To test the effectiveness of the Transductive Classifier, in this section, we compare HinDom with three inductive classification methods: Navie Bayes (NB), Support Vector Machine (SVM) and Random Forest (RF),on the situation where 90\%, 70\%, 50\%, 30\%, 10\% labels are kept randomly. For the inductive methods, we extract all details about entities and relations of the HIN instance as features to learn the classification functions. The \textit{accuracy} and \textit{F1 score} of each method are shown in Table 7. As we can see, HinDom maintains relatively stable performance when the fraction of initial label information decreases. It yields \textit{accuracy}: 0.9960, \textit{F1}:0.9905 when 90\% domains are pre-labeled and still obtains \textit{accuracy}: 0.9626, \textit{F1}: 0.9116 when only 10\% labels are left at the beginning. As for the inductive methods, they can obtain relatively good performance with sufficient labels, yet the accuracy drops to around 0.8 as the size of training dataset decreases. The reason behind is that by using HIN assisted with a transductive classifier, HinDom can not only learn from the labeled data but also fully exploit associations among domains and generally propagates the initial label information over the whole network. 

To find the minimum threshold of label fraction for relatively high performance, we gradually reduce the initial label information and draw curves of accuracy and F1 score in Figure 5. The breakpoint is around 10\%, which means we need to provide at least 10  percent labeled samples for domain detection, otherwise, the performance of HinDom will suffer a dramatic decrease.

\begin{figure}[hbtp]
	\centering
	\includegraphics[width =0.5\textwidth]{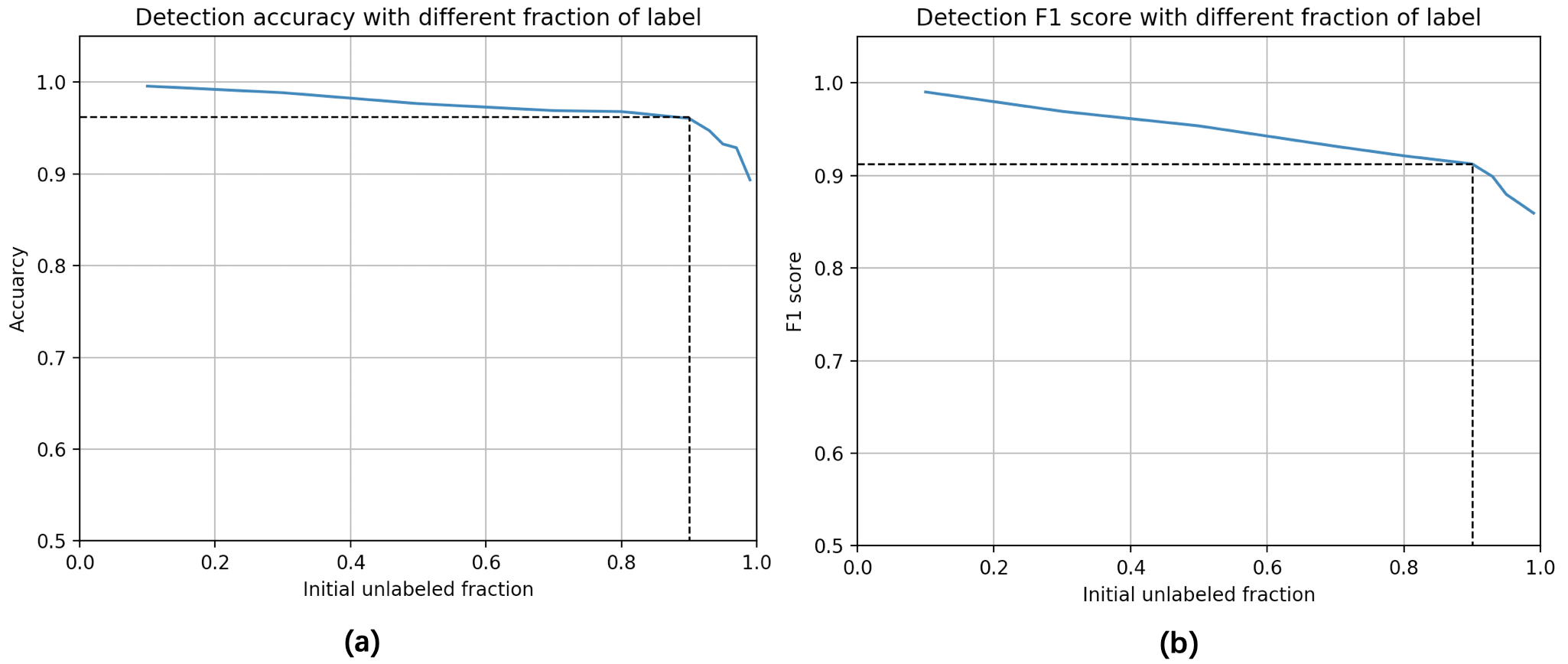}	
	\caption{Accuracy and F1 score with different initial label fraction}
\end{figure}

\subsection{Robutness}
Though consuming lots of efforts, the manually labeled training dataset often contains noises which might be caused by attackers' tricks or human mistakes. In this section, to test HinDom's robustness to label noises, we keep 70\% initial label information, randomly change labels of $k_d$\% training samples and compare the detection results of HinDom with those traditional methods: NB, SVM and RF. We increase $k_d$\% gradually and stop at 50\% where none of these methods can generate a tolerable detection result. We repeat each scenario for 10 times. Figure 6 shows the average accuracy trend and F1-score trend of each method. We can see that the traditional machine learning methods are susceptible to mislabeling. Meanwhile, with a better understanding of structural information, HinDom can hold relatively stable performance when dealing with some label noises.

\begin{figure}[hbtp]
	\centering
	\includegraphics[width =0.5\textwidth]{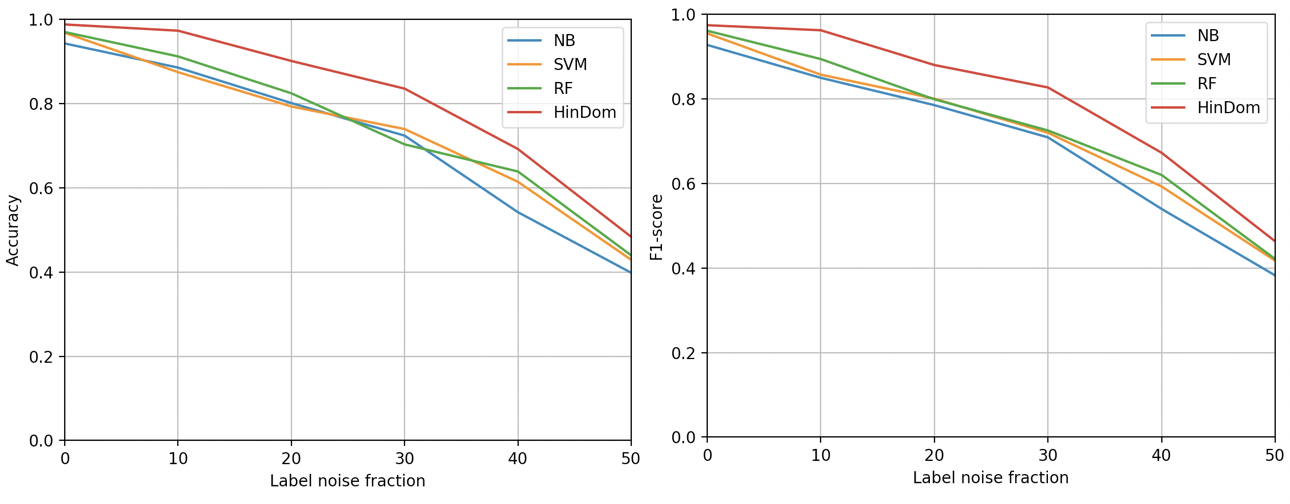}	
	\caption{Accuracy and F1 score with $k_d$\% label noise}
\end{figure}

\subsection{Multi-classification}

It is observed that malicious domains from the same attacker tend to follow the same patterns, regardless of character distributions, the victim groups or the sets of IP addresses they map to. From the definition of Transductive Classifier, we can see that apart from detecting malicious domains, HinDom can support multi-classification and identify which categories the malicious domains belong to. The domain family information provided by multi-classification is useful for follow-up work like reverse engineering and security reports. To test HinDom's multi-classification ability, we further label the 0.25 million malicious domains mentioned in Section 4.1 into 13 categories based on the malware or cybercrimines they related to. Note that for those categories with less than $F$=150 domains, we group their domains together as Class $Rare$. 

\begin{table}[htbp]
\small
\caption{Multi-classification with different fraction of labels}
\begin{center}
\begin{tabular}{|c|c|c|c|c|}
\hline
\textbf{Initial Label}&\multicolumn{4}{|c|}{Metrics}\\
\cline{2-5}
\textbf{Fraction}&\textbf{accuarcy}&\textbf{precision}&\textbf{recall}&\textbf{F1 Score}\\
\hline
90\%&0.9814&0.9786&0.9815&0.9801\\
\hline
70\%& 0.9783&0.9759&0.9765&0.9762\\
\hline
50\%&0.9720&0.9673&0.9706&0.9689\\
\hline
30\%&0.9644&0.96178&0.9654&0.9636\\
\hline
10\%&0.9598&0.9543&0.9585&0.95647\\
\hline
\end{tabular}
\label{tab1}
\end{center}
\end{table}

\begin{figure}[hbtp]
	\centering
	\includegraphics[width =0.45\textwidth]{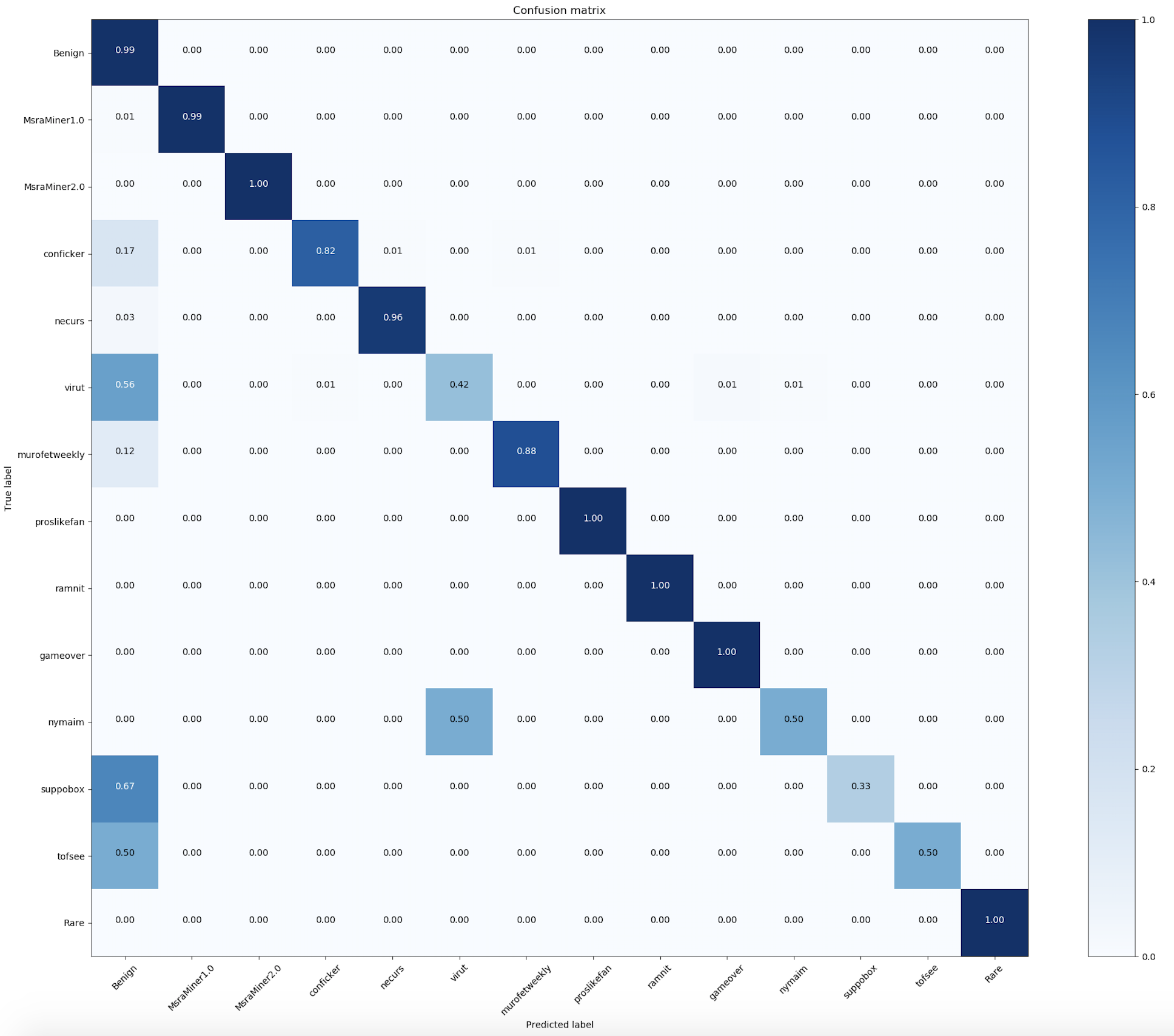}	
	\caption{Confusion matrix of multi-classification with 50\% initial labels}
\end{figure}

Table 8 lists the detection results with different fractions of initial labels. Considering the sample imbalance of each category, we use the weighted-average metrics to evaluate the multi-classification. With this method, the metrics of each label will be weighted by support to find their average. Due to space limitations, Figure 7 only displays the confusion matrix of multi-classification when there are 50\% initial labels. The X axis denotes the ture category of each domain while the Y axis denotes their predicted labels. The confuse matrix shows that most misclassifications are between Class \textit{Benign} and some malicious classes with relatively small sample size, which we suppose is caused by data skew. The additional information about domain family brings higher TPR yet lead to worse FPR. To solve this problem, we will separate HinDom into two stages, first distinguish malicious domains from the benign ones and then multi classify these malicious domains to identify their families.

\subsection{Public Information Only}

To test the real-world practicality,  we eliminate the influence of human decisions by running HinDom with initial labels only from public whitelists or blacklists. When using lists with only 2LDs (e.g. Alexa Top List, DGArchive), we set domains with the same 2LD to the same class. For instance, \textit{agoodm.m.taobao.com} and \textit{chat.im.taobao.com} are in benign class because their 2LD \textit{taobao.com} is in Alexa Top 1K. For those dubious domains appear in both whitelists and blacklists, we randomly set them to benign or malicious class. In the end, we label about 24\% of the 0.95 million domains with about 0.04 million confused ones. HinDom yields \textit{accuracy}: 0.9634, \textit{F1}: 0.9253 with this kind of initial label information.

Table 9 shows the performance o HinDom with public labels and with 20\%, 30\% manual labels while Figure 8 displays their ROC.
We can see that HinDom gets similar performance no matter with only public information or with manual labels of the same proportion. Besides, by analyzing the detection results, we find that with a relatively small $\mu=0.3$, which means we do not firmly insist on the pre-labeled information, HinDom can correct the label of some domains that are misclassified at the very beginning. For example, \textit{memberprod.alipay.com} and \textit{hosting.rediff.com} were randomly assigned to the malicious class because they have both benign and malicious information. HinDom adjusts their labels from 1 to 0 after 11 times iteration because of the strong associations they have with the benign domains.

\begin{table}[htbp]
\small
\caption{Detection results with public and manual labels}
\begin{center}
\begin{tabular}{|c|c|c|c|}
\hline
\textbf{Metrics}&\multicolumn{3}{|c|}{Labels}\\
\cline{2-4}
&\textbf{30\% Manual}&\textbf{Public only}&\textbf{20\% Manual}\\
\hline
\textbf{accuracy}&0.9698&0.9634&0.9633\\
\hline
\textbf{precision}&0.9510&0.9367&0.9380\\
\hline
\textbf{recall}&0.9396&0.9142&0.9087\\
\hline
\textbf{F1 score}&0.9453&0.9253&0.9232\\
\hline
\end{tabular}
\label{tab1}
\end{center}
\end{table}

\begin{figure}[hbtp]
	\centering
	\includegraphics[width =0.45\textwidth]{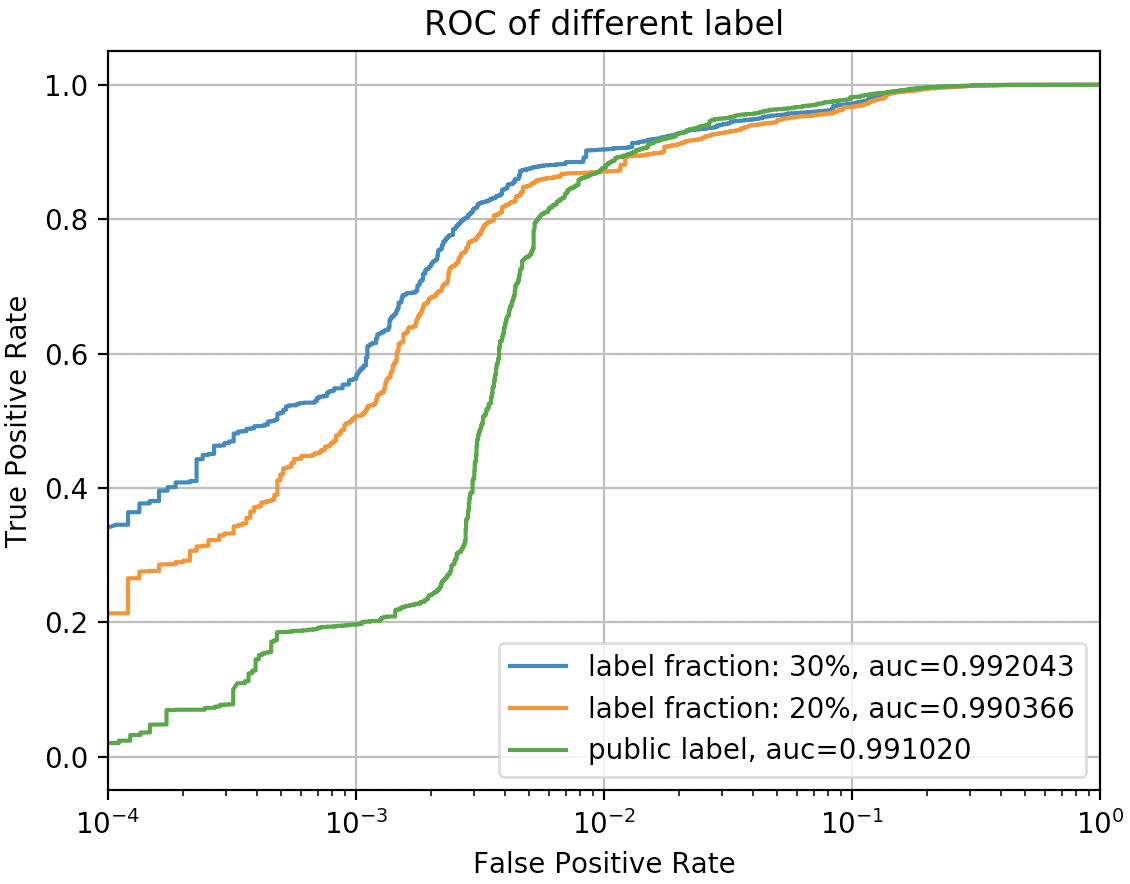}	
	\caption{ROC of detection with public or manual labels}
\end{figure}

\subsection{Compare with detection engines}
To test HinDom's practicality, we also compare its performance with existing Detection Engines (DE) on VirusTotal\cite{Virustotal}. We randomly choose 30\% domains from the experimental dataset as testing samples and collect the detection results of all 66 engines in VirusTotal. Due to space limitations, Table 10 only displays nine engines that have relatively good performance or that are widely used by security vendors like McAfee, Kaspersky, etc. We can see that HinDom outperforms six out of nine engines with \textit{accuracy}: 0.9697 and \textit{F1-score}: 0.9460. This is a remarkable result as these engines are supported by more expertise and the label information of our experimental dataset is partly depended on VirusTotal.

\begin{table}[htbp]
\small
\caption{Detection results of engines from VirusTotal}
\begin{center}
\begin{tabular}{|c|c|c|c|c|c|}
\hline
\textbf{Engines}&\textbf{HinDom}&\textbf{DE1}&\textbf{DE2}&\textbf{DE3}&\textbf{DE4}\\
\hline
ACC&0.9697&0.9038&0.9785&0.8952&0.9603\\
\hline
F1-score&0.9460&0.8898&0.9701&0.8824&0.9511\\
\hline
\textbf{Engines}&\textbf{DE5}&\textbf{DE6}&\textbf{DE7}&\textbf{DE8}&\textbf{DE9}\\
\hline
ACC&0.9812&0.9846&0.9023&0.9116&0.8974\\
\hline
F1-score&0.9747&0.9800&0.8975&0.9048&0.8423\\
\hline
\end{tabular}
\label{tab1}
\end{center}
\end{table}

\subsection{Real-world Deployment}

We implement a prototype system of HinDom and deploy it in CERNET2 and TUNET. Though HinDom is not a real-time detection system, it can be run every hour or every day to detect ongoing malicious activities. Taking computing resources and the network scale into consideration, we choose to set the time window \textit{T} to one hour in our deployment. In other words, we construct a HIN instance based on the DNS-related data within an hour, get the detection results, accumulate label information, and the cycle repeats.

In CERNET2, on average, about 40 thousand clients initiate 3.8 million DNS requests within an hour and about 0.25 million unique domains exist after graph pruning. With HinDom we find 3.34\% domains are malicious. The result is proved to be reliable by experts certification and some of these malicious domains are detected several months before they are reported by public services. In particular, HinDom detects a bunch of domains that are not listed in any public blacklists,  neither are the IP addresses they are resolved to. After consulting Qihoo 360, a Chinese internet security company, we confirm these domains belong to a long-buried mining botnet named \textit{MsraMiner}. In TUNET, about 50 thousand clients request for 0.4 million unique domains per hour. With closer network supervision, the proportion of malicious domains drops to 1.21\%. However, we still detect the variation of \textit{MsraMiner}, with domains like \textit{ra1.kziu0tpofwf.club} and \textit{sim.jiovt.com}, in the campus network. The above detection results have been reported to the relevant network management department.

\section{Limitation and Future Work}

Currently, HinDom constructs a HIN of clients, domains and IP addresses based on the DNS-related data and can perform well even with a small fraction of labeled samples. Yet there are some potential problems in views of scalability and practicability. We discuss HinDom's limits and our future work in this section.

First, efficiency. With a graph-based mechanism, HinDom cannot be deployed in a real-time mode. We need to choose a proper time window $T$ to collect data and then conduct the detection procedure off-line. If $T$ is too small, the collected data will not be sufficient for accurate detection. Yet with a very big time window, HinDom will need more computing resources and longer detection time. Thus, there is a trade-off between accuracy and efficiency. We have mentioned that the multiplication between adjacency matrices in HinDom is a resource-consuming operation. We recommend matrix block calculation and parallel computation framework like Hadoop in the real-world deployment. Besides, we find that embedding is an up-and-coming approach to represent graphs in a low dimensional way. Some embedding methods such as HIN2Vec\cite{fu2017hin2vec} and ESim\cite{shang2016meta} have been proposed to represent nodes in HIN, which can be used to improve HinDom's efficiency.

Second, the detection range. As an association-based detection system, HinDom can only detect malicious domains that have direct or indirect relations with others. When a new type of malicious domain is just registered and has few relations with other entities, HinDom cannot detect it immediately. Besides, HinDom may not hold a good performance when detecting malicious domains hosted by network services like CDN, as they are related to a large number of benign domains. To eliminate these problems, we plan to utilize more types of DNS-related data and dig out richer associations among domains. For example, WHOIS dataset is an important clue with information about registered users or assignees. Besides, for now, we only use A, AAAA and CNAME records in DNS traffic and may expand to PDG, MX, SRV, NS, PTR for richer information in the future.

Third, further analyzation. After getting domain detection results, we can design functions and go further to find the infected clients and malicious IPs in the network based on the rich semantic information represented in HinDom. With this information, the security management team can narrow down their investigation range and focus on the most dangerous hosts. We will add the client and IP detection module in future work.

\section{Conclusion}

In this paper, we present an intelligent malicious domain detection system named HinDom. HinDom constructs a HIN of clients, domains and IP addresses to model the DNS scene and generates a combined meta-path to analyze the associations among domains. With a meta-path based transductive classification method, HinDom performs well even when the initial label fraction drops to 10\%, which reduces the cost of acquiring labeled samples. In our extensive evaluation, we verified HinDom’s performance, robustness and practicality. In the real-world deployment, we are able to discover a long-buried mining botnet named MsraMiner and some other malicious domains ahead of the public services. As for further developments, we plan to extend to other types of DNS-related data for more comprehensive semantic information and integrate graph embedding in HinDom to improve efficiency.

\section*{Acknowledgment}
We thank Hui Zhang, Chenxi Li, Shize Zhang for constructive recommendations on experiments and data processing. Additionally, we appreciate 360netLab, VirusTotal for permissions of their advanced APIs and we thank Information Technology Center of Tsinghua University for authorizing the use of their data in our experiments. This work is supported by the National Key Research and Development Program of China under Grant No.2017YFB0803004.

\bibliographystyle{plain}
\bibliography{hindomref}

\end{document}